\documentclass[prb,twocolumn,showpacs]{revtex4}

\usepackage{graphicx}
\usepackage{subfigure}
\usepackage{dcolumn}
\usepackage{bm}
\usepackage{multirow}

\def\ket#1{\mathinner{|{#1}\rangle}}

\def\beq{\begin{equation}}
\def\eeq{\end{equation}}

\begin{document}
\author{Avinash Kolli}
\email{avinash.kolli@materials.ox.ac.uk}
\affiliation{Department of Materials, Oxford University, Oxford OX1 3PH, UK}
\author{Simon C. Benjamin}
\affiliation{Department of Materials, Oxford University, Oxford OX1 3PH, UK}
\author{Brendon W. Lovett}
\affiliation{Department of Materials, Oxford University, Oxford OX1 3PH, UK}
\author{Thomas M. Stace}
\affiliation{Department of Physics, University of Queensland, Brisbane, QLD, Australia}

\title{Measurement-based approach to entanglement generation in coupled quantum dots}

\begin{abstract}
Measurements provide a novel mechanism for generating the entanglement resource necessary for performing scalable quantum computation. Recently, we proposed a method for performing parity measurements in a coupled quantum dot system~\cite{kolli06}. In this paper we generalise this scheme and perform a comprehensive analytic and numerical study of environmental factors. We calculate the effects of possible error sources including non-ideal photon detectors, ineffective spin-selective excitation and dot distinguishability (both spatial and spectral). Furthermore, we present an experimental approach for verifying the success of the parity measurement.
\end{abstract}

\maketitle

\section{Introduction}


Quantum dots are semiconductor hetero-structures which exhibit strong electron and hole confinement. This leads to a highly discrete level structure, giving rise to many interesting nanotechnological applications, such as quantum information processing (QIP). Such a discrete level structure enables us to identify well-defined effective two-level systems (qubits), which we use to encode our quantum information. A natural qubit is the spin of an excess electron within a quantum dot, which typically exhibits long lifetime (up to milliseconds~\cite{kroutvar04, elzerman04}) and coherence times (up to microseconds~\cite{petta05}).

For universal quantum computation (QC) we must be able to perform any one qubit gate, plus an entangling two qubit gate. Typically we utilise the natural couplings between the spins to generate our two qubit gate. Experimentally these two qubit gates would be achieved as periods of free evolution of the interacting system followed by periods of controlled single qubit rotations. For example, early proposals \cite{loss98} included using the the exchange interaction $H = J \hat{S}_{1} \cdot \hat{S}_{2}$ to provide the necessary two qubit interaction. However, precise manipulation of the interactions between the spins is needed for such a scheme to work, which might be difficult to achieve in practice.



An alternative method, which we discuss here, is to utilise the power of measurements to generate interactions between the qubits. Such measurement-based ideas initially appeared in the context of linear optical approaches to quantum computation~\cite{knill00}.
In 2000, Knill, Laflamme and Milburn proposed a method for generating entanglement using only linear optical elements, ancilla photons and measurements on single photons.
A number of proposals have since been presented for generating entanglement using optics in solid-state systems. For example, Refs.~\onlinecite{lim05} and \onlinecite{barrett05a}  propose using single photon interference to entangle spatially separated matter qubits.

If we could implement a measurement-based scheme with the electron spin we could do away with the exquisite control required for the conventional approaches discussed above. However, there exists a no-go theorem which states that it is not possible to achieve an exponential speed-up over classical computation using solely single electron Hamiltonians and single spin-measurements \cite{terhal02}. Recently, Beenakker {\it et al.}~\cite{beenakker04} have shown that it is possible to lift this restriction if we look outside the Hilbert space of a spin and exploit the charge degree  of freedom. Charge and spin commute, and so we are able to make measurements on the charge without destroying any information that is contained in the spin degrees of freedom. Beenakker {\it et al.} proceed to show that partial-Bell state measurements (also known as parity measurements) on the spin states are sufficient to implement a CNOT gate, thus enabling universal QC. A number of subsequent papers have proposed specific implementations for these spin-parity measurements. These include a charge tunneling detection method~\cite{engel05}), and a charge fluctuation method~\cite{barrett06b}.


Recently, we proposed a method for optically performing parity measurements on a pair of coupled quantum dots. In this paper we generalise this scheme and perform a comprehensive study of environmental factors. We will begin, in section \ref{sec:model} by outlining the system and the interactions present. In section \ref{sec:scheme} we will detail the various steps in the spin-parity measurement. We will present results for the operation of the measurement in section \ref{sec:results} , and then in section \ref{sec:errors} we will analyze the effect of various error mechanisms, including valence band mixing, and spatial and spectral distinguishability of the two dots. We then proceed to describe a method for verifying the success of our gate in section~\ref{sec:verification} . Finally, we will give some concluding remarks.

\newpage

\section{The Coupled Quantum Dot System}


\label{sec:model}

\begin{figure}[t]
\includegraphics[width = 0.45\textwidth]{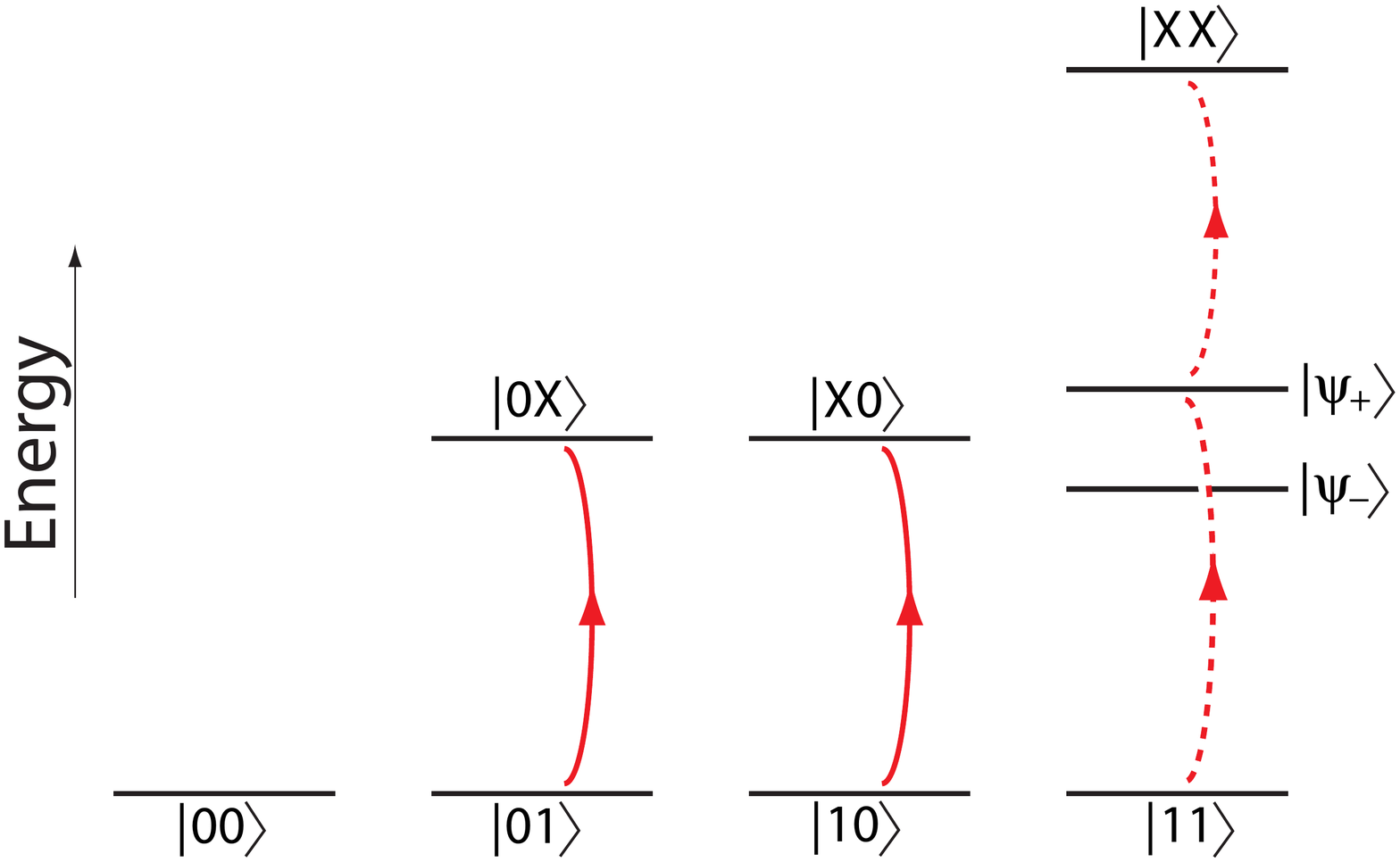}
\caption{Energy level diagram for the coupled quantum dot structure. Resonant transitions are denoted by the solid lines, while non-resonant transitions by dotted lines.}
\label{fig:levels}
\end{figure}


In order to develop our coupled model, we will first introduce the structure of a single quantum dot interacting with a laser field. We assume that the dot is $n$-doped such that the valence levels (VL) are completely filled and the only occupied conduction level (CL) is the lowest lying level. In what follows, we will only consider the top most filled valence states $|J_z = \pm 3/2\rangle$. The CL electron has a spin degree of freedom on which we encode our quantum information: $|0\rangle$ is encoded in the $m_{z}=-1/2$ state and $|1\rangle$ in $m_{z}=1/2$. The system is irradiated by a classical $\sigma^+$ circularly polarised
laser field, resonant with the VL-CL energy gap. An electron-hole pair (exciton) state can be created, if a photon's angular momentum of $+\hbar$ can be absorbed; Pauli's exclusion principle means this is only possible when the qubit is in state $|m_{z}=-1/2\rangle =|1\rangle$.
This effect is `Pauli Blocking', and it enables us to generate excitons conditioned on the state of the qubit electron. The combined qubit electron/exciton state is a trion, denoted by $|X\rangle$.

When two such doped dots are placed close to each other, there are direct electron spin-spin couplings. However, these are very weak: current experiments place their strengths at less than $1~\mu eV$~\cite{laird06}. We take these interactions to be negligible in comparison to the excitonic interaction that we will exploit. The two primary excitonic couplings that we consider are a static and a dynamic dipole-dipole coupling. The static interaction results in an energy shift $V_{XX}$ of the double trion state and is due to the inter-dot exciton-exciton dipole interaction. The dynamic coupling, or resonant Foerster interaction $V_F$, results in an exciton transfer from one dot to the other that is mediation by a virtual photon. This process has been shown, to first-order, to be non-magnetic and thus conserves electron spin~\cite{lovett05}. Therefore, in this system the Foerster interaction only couples the $|X1\rangle$ and $|1X\rangle$ states. The Hamiltonian for the coupled QDs is:

\begin{eqnarray}
\label{eq:H} H &=& \omega_{0} |X\rangle \langle X| \otimes \hat{I} + \omega_{0} \hat{I} \otimes |X\rangle \langle X| +
\nonumber\\ && V_{XX} |XX\rangle \langle XX| + V_{F}\big(|1X\rangle \langle X1| + H.c.\big) \nonumber\\ && + \Omega \cos \omega_{l}t \big(|1\rangle \langle X| \otimes
\hat{I} + \hat{I} \otimes |1\rangle \langle X| + H.c.\big),
\end{eqnarray}

\noindent where $H.c.$ denotes Hermitian conjugate, $\omega_{0}$ is the exciton creation energy for both dots (the dots are assumed to be identical), $\Omega$ is the time-independent laser coupling (assumed to be the same for both dots), and $\omega_{l}$ is the laser frequency. The energy difference between the $|0\rangle$ and $|1\rangle$ states is negligible on the exciton energy scale.

The Hamiltonian (\ref{eq:H}) may be decoupled into four subspaces with no interactions between them: $H_{00} = \{ |00\rangle \}$, $H_{01}=\{ |01\rangle ,|0X\rangle \}$ , $H_{10} = \{ |10\rangle , |X0\rangle \}$, $H_{11}=\{ |11\rangle , |X1\rangle, |X1\rangle , |XX\rangle \}$. Let us look more closely at the Hamiltonian for the last of these subspaces. We rewrite this in basis of the eigenstates when $\Omega = 0$, which are $|11\rangle$, $|\psi_{+}\rangle = \frac{1}{\sqrt{2}}(|1X\rangle + |X1\rangle)$, $|\psi_{-}\rangle =\frac{1}{\sqrt{2}} (|1X\rangle - |X1\rangle)$ and $|XX\rangle$. The degeneracy of the $|\psi_{-}\rangle$ and $|\psi_{+}\rangle$ levels is lifted by the Foerster interaction, resulting in two states each containing a delocalized exciton. In this basis the Hamiltonian is, after re-introducing a finite $\Omega$:

\begin{eqnarray} \label{eq:H2} H_{11} &=& (\omega_{0} + V_{F}) |\psi_{+}\rangle \langle\psi_{+}| + (\omega_{0} - V_{F}) |\psi_{-}\rangle \langle \psi_{-}| \nonumber\\ && +
(2 \omega_{0} + V_{XX})|XX\rangle \langle XX| \nonumber\\ && + \Omega^{'} \cos \omega_{l}t \big(|11\rangle \langle \psi_{+}| + |\psi_{+}\rangle \langle XX| + H.c. \big) .
\end{eqnarray}

\noindent The only dipole allowed transitions in this subspace are between $|11\rangle$ and $|\psi_{+}\rangle$, and between $|\psi_{+}\rangle$ and
$|XX\rangle$, with a coupling strength of $\Omega^{'}=\sqrt{2} \Omega$. The level structure for all four subspaces is shown in Fig. \ref{fig:levels}.



\section{Spin Parity Measurement}


\label{sec:scheme}

The parity measurement protocol consists of two steps: excitation followed by monitored relaxation. In the excitation we aim to transfer the population of the states from the computational basis into the excitonic levels in the odd-parity  ($\ket{10}-\ket{01}$) subspace, while retaining the population of the even-parity  ($\ket{00}-\ket{11}$) states in the ground levels. We can achieve this by exciting the coupled dots with a pulsed laser tuned to energy $\omega_0$. By referring to Fig.~\ref{fig:levels} and Eqs.~\ref{eq:H2} and \ref{eq:H}, we can see that such a laser excites transitions within  the $\ket{01}$ and $\ket{10}$ subspaces. The $\ket{11}$ state is not excited to first order if
\begin{equation}
\label{eq:cond} |V_{F}|, |V_{XX}| \gg |\Omega^{'}|/2,
\end{equation}
since the laser is off resonance with the excited levels in this space, and $\ket{00}$ is of course optically inactive.

Suppose we have an initial state
\beq
\label{initialstate}
|\psi_{0}\rangle = \alpha_{00} |00\rangle + \alpha_{01} |01\rangle +\alpha_{10} |10\rangle +\alpha_{11} |11\rangle.
\eeq
After an excitation $\pi$ pulse, our state is $|\psi\rangle = \alpha_{00} |00\rangle + \alpha_{01} |0X\rangle +\alpha_{10} |X0\rangle +\alpha_{11} |11\rangle$. We next enter a period of monitoring the system for decay photons. Assuming perfect detection, the state of the system is projected into the odd parity subspace if a photon is detected, and into the even parity subspace if a photon is not detected. Importantly, we only distinguish between the two parity subspaces without distinguishing states within the same subspace. We can represent the action of the measurement in terms of projection operators for the two desired outcomes:

\begin{eqnarray}
P_{O} &=& |\psi_{odd}\rangle\langle \psi_{odd}| = |01\rangle\langle 01| + |10\rangle\langle 10| , \nonumber\\
P_{E} &=& |\psi_{even}\rangle\langle \psi_{even}| = |00\rangle\langle 00| + |11\rangle\langle 11|.
\end{eqnarray}

As mentioned earlier, we must not be able to distinguish between states within the same subspace. Therefore it is important throughout the radiative relaxation that there is no information gained about the source of the photon that is emitted. There are many ways in which this condition can be compromised: for example spatial and spectral distinguishability of photon emissions from the different dots. Our system will also have imperfect detectors; in the next section we will begin to look at these potential sources of error.

\section{Conditional Dynamics with Imperfect Detectors}


\label{sec:results}

To model the monitored radiative relaxation we use the quantum trajectories formalism \cite{wiseman94,gardiner00,stace03}. The conditional master equation (CME) describing $n$ monitored relaxation channels is:

\begin{widetext}
\begin{eqnarray}
\label{eq:CME} d\rho_{c} &=& -i[H,\rho_{c}]dt + \sum_{j}^{n} \Big\{ \eta_j Tr ( \mathcal{J}[c_j] \rho_{c}) \rho_{c} + (1-\eta_j) \mathcal{J}[c_j] \rho_{c} - \mathcal{A}[c_j]\rho_{c} \Big\} dt + \Big\{ \frac{ \mathcal{J}[c_j] \rho_{c} }{Tr(\mathcal{J}[c_j] \rho_{c}) } - \rho_{c} \Big\} dN_j(t)
\end{eqnarray}
\end{widetext}

\noindent where $\rho_{c}$ is the density matrix of the system, $H$ is the system Hamiltonian in the interaction picture, $c_j$ is the Lindblad operator through which the system couples to the measurement channel $j$, $\mathcal{J}[c_j]$ is the jump super-operator which projects out the component of the state that is consistent with a detection from channel $j$ and is defined as $\mathcal{J}[c_j] \rho_c =c^{\dagger}_j \rho_c c_j$. $\mathcal{A}[c_j]$ is defined to be $\mathcal{A}[c_j]\rho_c=\frac{1}{2} (c^{\dagger}_j c_j \rho_c + \rho_c c^{\dagger}_j c_j)$; $\eta_j$ is the efficiency of the detector that monitors emission into channel $j$. $dN_j(t)$ is the classical stochastic increment taking the values $\{0,1\}$, which denotes the number of photons detected in channel $j$ in the interval $t , t+dt$.

Between quantum jumps, when $dN(t) = 0$, Eq.~\ref{eq:CME} is equivalent to the linear, unnormalised, CME

\beq
\label{UNcme}
\dot{\tilde{\rho}} = - i [H,\tilde{\rho}] + \sum_{j}^{n} \big\{(1-\eta_{j}) \mathcal{J}[c_j]\tilde{\rho} - \mathcal{A}[c_j]\tilde{\rho}\big\} \end{equation}

\noindent where $\rho_c = \tilde{\rho}/Tr(\tilde{\rho})$.

Assuming that the condition (\ref{eq:cond}) is satisfied, there are no excitons in the even-subspace. Thus we need only consider one channel defined by the Lindblad operator $c = \sqrt{\Gamma_{X}} (|01\rangle \langle0X| + |10\rangle \langle X0|)$. Furthermore, the odd-space only contains a single excitation and thus we can model the dynamics in two steps: a period of continuous evolution followed by a potential single quantum jump due to the photon detection event.

Suitable parameters are chosen and are displayed in the Table~\ref{params}~\cite{birkedal01,biolatti02,lovett03b,borri01}:

\begin{center}
\begin{table}
\label{params}
\begin{tabular}{c c}
\hline
Parameters &
Value \\
\hline
$V_{F}$ & $0.85$~meV \\
$V_{XX}$ & $5$~meV \\
$\omega_{0}$ & $2$~eV \\
$\Omega$ & $0.1$~meV \\
$\tau_{X}$ & $1$~ns \\
$\Gamma_{X}$ & $4 ~\mu$eV \\
\hline
\end{tabular}
\caption{Table of relevant parameters for the coupled quantum dot system}
\end{table}
\end{center}

For an initial state (\ref{initialstate}), and using Eq.~\ref{UNcme}, we find that the probability $p_{E}$ that we are in the even parity subspace at a time $t$ after excitation is:

\begin{equation}
\label{pE}
p_{E}(t) = \frac{\alpha_{00}^2 + \alpha_{11}^2}{1 + \eta (\alpha_{01}^2 + \alpha_{10}^2) (e^{-\Gamma_{X} t} - 1)}.
\end{equation}

In order to increase this probability, it makes sense to wait for long enough that we can be sure that if a photon has not yet been emitted, it is not likely to be emitted in future. This amounts to waiting for a time $t\gg 1/\Gamma_X$.
Then the fidelity of projection into the even-subspace is:
\begin{equation}
F_{E} = \frac{\alpha_{00}^2 + \alpha_{11}^2}{(1 - \eta)(\alpha_{01}^2 + \alpha_{10}^2) + (\alpha_{00}^2 + \alpha_{11}^2)}.
\end{equation}
When a photon is detected (and ignoring typically negligible detector dark counts) the fidelity of for projection into the odd-subspace is $F_{O} = 1$ .

The success of our two step parity measurement is strongly dependent on detector efficiency, and can become quite poor for typical values of $\eta$. However, by repeating the spin-parity measurement one or more times, it is possible to obtain improved fidelities. On each round of the repeated measurement, we gain greater confidence that we have successfully projected into the even subspace, rather than missing every emitted photon.
To analyze this we write the effect of the spin parity measurement when no photon is measured in the quantum operation formalism \cite{nielsen00}. The action of a general quantum operation can be written as:
\beq
\rho \rightarrow \mathcal{E}(\rho) = \sum_{k} \tilde{E}_{k} \rho \tilde{E}_{k}^{\dagger},
\eeq
where the $\tilde{E}_{k}$ are the Krauss projection operators which must satisfy the normalistion condition $\sum_{k} \tilde{E}_{k}^{\dagger} \tilde{E}_{k} = I$. For our particular example, a single operation of the spin parity measurement will yield:
\beq
\rho \rightarrow \mathcal{E}(\rho) = \frac{c_{O} \tilde{E}_{O} \rho \tilde{E}_{O}^{\dagger} + c_{E} \tilde{E}_{E} \rho \tilde{E}_{E}^{\dagger}}{Tr[c_{O} \tilde{E}_{O} \rho \tilde{E}_{O}^{\dagger} + c_{E} \tilde{E}_{E} \rho \tilde{E}_{E}^{\dagger}]},
\eeq
where the un-normalised Krauss operators, defined as $\tilde{E}_{O,E} = P_{O,E} = |\psi_{O,E}\rangle\langle\psi_{O,E}|$, are the projectors onto the odd and even parity subspaces respectively, and the coefficients are chosen in a self-consistent manner: $c_{E} = 1$ and $c_{O} = (1-\eta)$. The denominator in this expression ensures that we satisfy the normalisation condition.

When the measurement is repeated $r$ times, the overall quantum operation is given by:
\beq \rho \rightarrow \mathcal{E}(\mathcal{E}(\rho)) = \frac{c^{r}_{O} \tilde{E}_{O} \rho \tilde{E}_{O}^{\dagger} + c^{r}_{E} \tilde{E}_{E} \rho \tilde{E}_{E}^{\dagger}}{Tr[c^{r}_{O} \tilde{E}_{O} \rho \tilde{E}_{O}^{\dagger} + c^{r}_{E} \tilde{E}_{E} \rho \tilde{E}_{E}^{\dagger}]}.
\eeq
The fidelity of correctly projecting into the even subspace is therefore given by:
\begin{eqnarray}
F_{E}^{r} &=& \frac{Tr[c^{r}_{E} \tilde{E}_{E} \rho \tilde{E}_{E}^{\dagger}]}{Tr[c^{r}_{O} \tilde{E}_{O} \rho \tilde{E}_{O}^{\dagger} + c^{r}_{E} \tilde{E}_{E} \rho \tilde{E}_{E}^{\dagger}]} \nonumber \\
&=& \frac{(\alpha_{00}^2+\alpha_{11}^2)}{(1-\eta)^{r}(\alpha_{01}^2+\alpha_{10}^2) + (\alpha_{00}^2+\alpha_{11}^2)}. \nonumber \\
\end{eqnarray}

For all non-zero detector efficiencies, in the limit $r\rightarrow\infty$ the term $(1-\eta)^r$ will tend to zero. We can therefore expect a unit fidelity for every input state in this limit.

Let us now look at the average fidelity for all input states as a function of $r$. Owing to the normalisation condition $\alpha_{00}^2 + \alpha_{01}^2 + \alpha_{10}^2 + \alpha_{11}^2 = 1$, we write the four coefficents in terms of four-dimensional hyperspherical polar coordinates:
\begin{eqnarray}
\alpha_{00} &=& \sin \phi_{1} \sin \phi_{2} \cos \phi_{3} \nonumber \\
\alpha_{01} &=& \cos \phi_{1}  \nonumber \\
\alpha_{10} &=& \sin \phi_{1} \cos \phi_{2}  \nonumber \\
\alpha_{11} &=& \sin \phi_{1} \sin \phi_{2} \sin \phi_{3} \nonumber \\
\end{eqnarray}
and the area element is given by:
\begin{equation}
d A = \sin^{2} \phi_{1} \sin \phi_{2} \, d \phi_{1} \, d \phi_{2} \, d \phi_{3}
\end{equation}
The resulting integral is:
\begin{equation}
\bar{F}_{E}^{r} = \int_{0}^{\pi} d \phi_{1} \sin^{2} \phi_{1} \int_{0}^{\pi} d \phi_{2} \sin \phi_{2} \int_{0}^{2\pi} d\phi_{3} \, F_{E}^{r}.
\end{equation}

\begin{figure}[t]
\label{repeat}
\includegraphics[width = 0.4\textwidth]{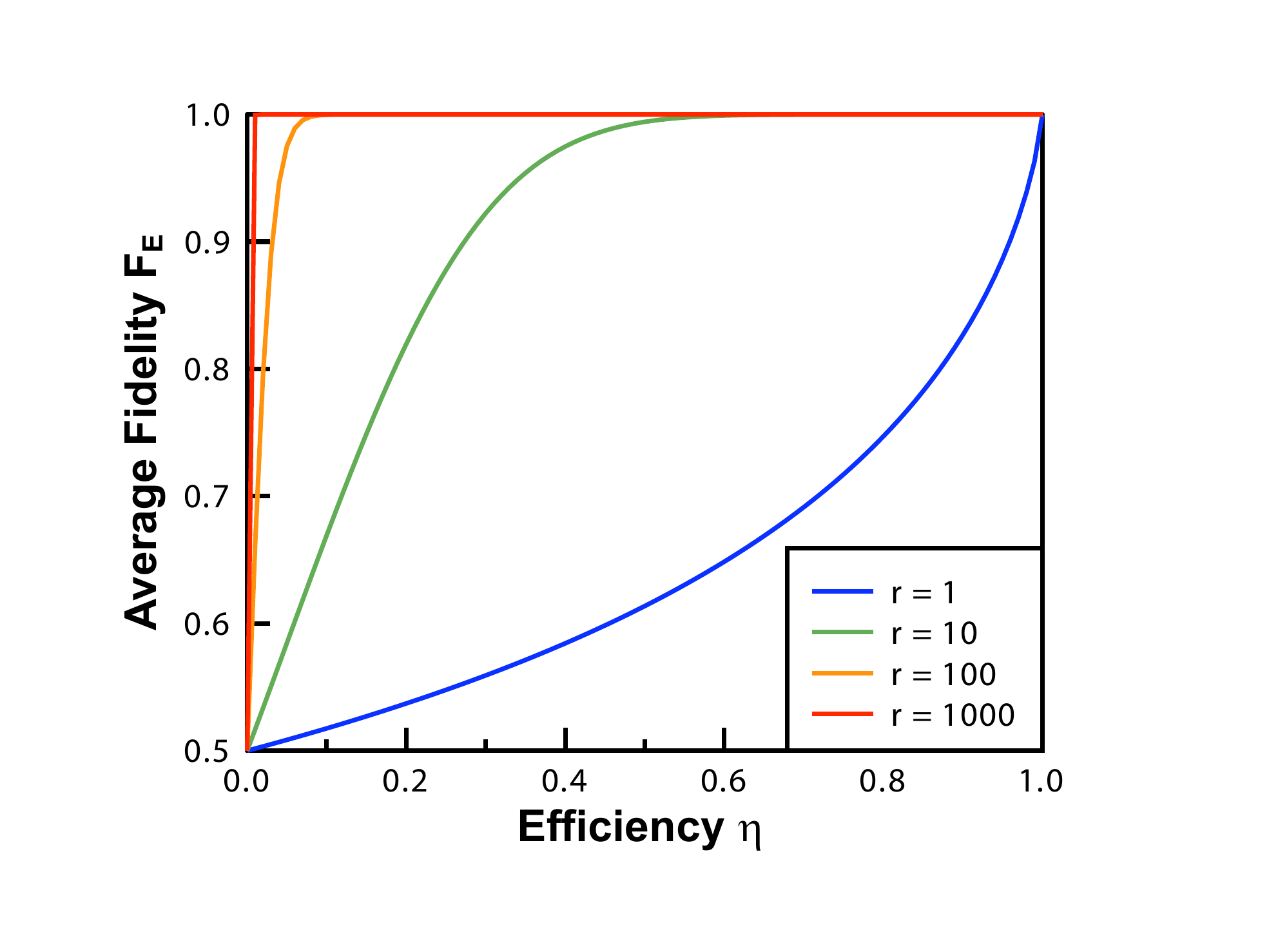}
\caption{The averaged fidelities $\bar{F}_{E}$ as a function of detector efficiency.}
\label{fig:repeat}
\end{figure}

This integration is performed numerically and the resulting averaged fidelity as a function of $\eta$ can be seen in Fig. \ref{fig:repeat}. We clearly see a convergence, as $r\rightarrow\infty$, of the average fidelity to unity for all non-zero detection efficiencies. Thus, by simply repeating the measurement, we are able to overcome the inherent problems of loss detectors.

\section{Corrections to Model}

\label{sec:errors}

\subsection{Valence band mixing}


\label{sec:vbmixing}

The Pauli blocking mechanism, crucial to the success of the excitation step, is valid only in the case of no light-heavy hole mixing, which is only true in very limited cases~\cite{bayer99}. Usually the hole eigenstates are composed of mixtures of light hole ($|J_z = \pm 1/2\rangle$) and heavy hole ($|J_z = \pm 3/2\rangle$) states~\cite{luttinger55}. The mixing is characterised by a factor $\epsilon$. This results in two families of eigenstates: one with predominantly light hole character and one with predominantly heavy hole character. The latter tend to be the topmost valence levels and are~\cite{lovett05}:
\begin{eqnarray}
|h_{+}\rangle &=& \sqrt{1-\epsilon^2}|J_z = +3/2\rangle + \epsilon |J_z = -1/2\rangle \nonumber \\
|h_{-}\rangle &=& \sqrt{1-\epsilon^2}|J_z = -3/2\rangle + \epsilon |J_z = +1/2\rangle.
\end{eqnarray}
On applying the $\sigma^{+}$ polarised laser field, it is now possible to generate excitons from both the $|0\rangle = |m_z = -1/2\rangle$ and $|1\rangle = |m_z = +1/2\rangle$ electron states. The resulting exciton-laser field coupling Hamiltonian is:
\begin{equation}
H_{\sigma^{+}} = \cos[\omega_{L}t] (|1\rangle\langle X_{-}| + \tilde{\epsilon} |0\rangle\langle X_{+}| + H. c.),
\end{equation}
where the trion levels are $|X_{+,-}\rangle  = |S_{\uparrow\downarrow}\rangle \otimes |h_{+,-}\rangle$. The modified mixing angle is $\tilde{\epsilon} = \epsilon \frac{l_{lh}}{\sqrt{3} l_{hh}}$ and $l_{lh,hh}$ are characteristic lengths associated with the overlap of the electron and the ${lh,hh}$ Bloch functions (see Ref.~\onlinecite{lovett05}).

The mixing also induces a Foerster interaction that couple other single exciton levels:
\begin{eqnarray}
H_{F} &=& M_{hh,hh} (|0X_{+}\rangle\langle X_{+} 0| + |1X_{-}\rangle\langle X_{-}1|) + \nonumber \\
&& \frac{2M_{lh,hh} \epsilon}{\sqrt{3}} (|0X_{-}\rangle\langle X_{+}1| + |1X_{+}\rangle\langle X_{-}0|) + h.c. \nonumber\\
\end{eqnarray}

\noindent where $M_{i,j}$ are the matrix elements for the transitions induced by the Foerster interaction and $i$ and $j$ denote the different initial and final hole states respectively.

The Hamiltonian for the coupled quantum dots is now:
\begin{eqnarray}
H &=& \omega_{0} (|X_{+}\rangle\langle X_{+}| + |X_{-}\rangle\langle X_{-}|)\otimes I + \nonumber \\ && \omega_{0} I \otimes (|X_{+}\rangle\langle X_{+}| + |X_{-}\rangle\langle X_{-}|) \nonumber \\ && + H_{F} + H_{\sigma^{+}} + \sum_{\mu,\nu \in \{X_{-},X_{+}\}} V_{XX} |\mu \nu\rangle\langle \mu\nu|
\end{eqnarray}

\begin{figure}[t]
\includegraphics[width = 0.5\textwidth]{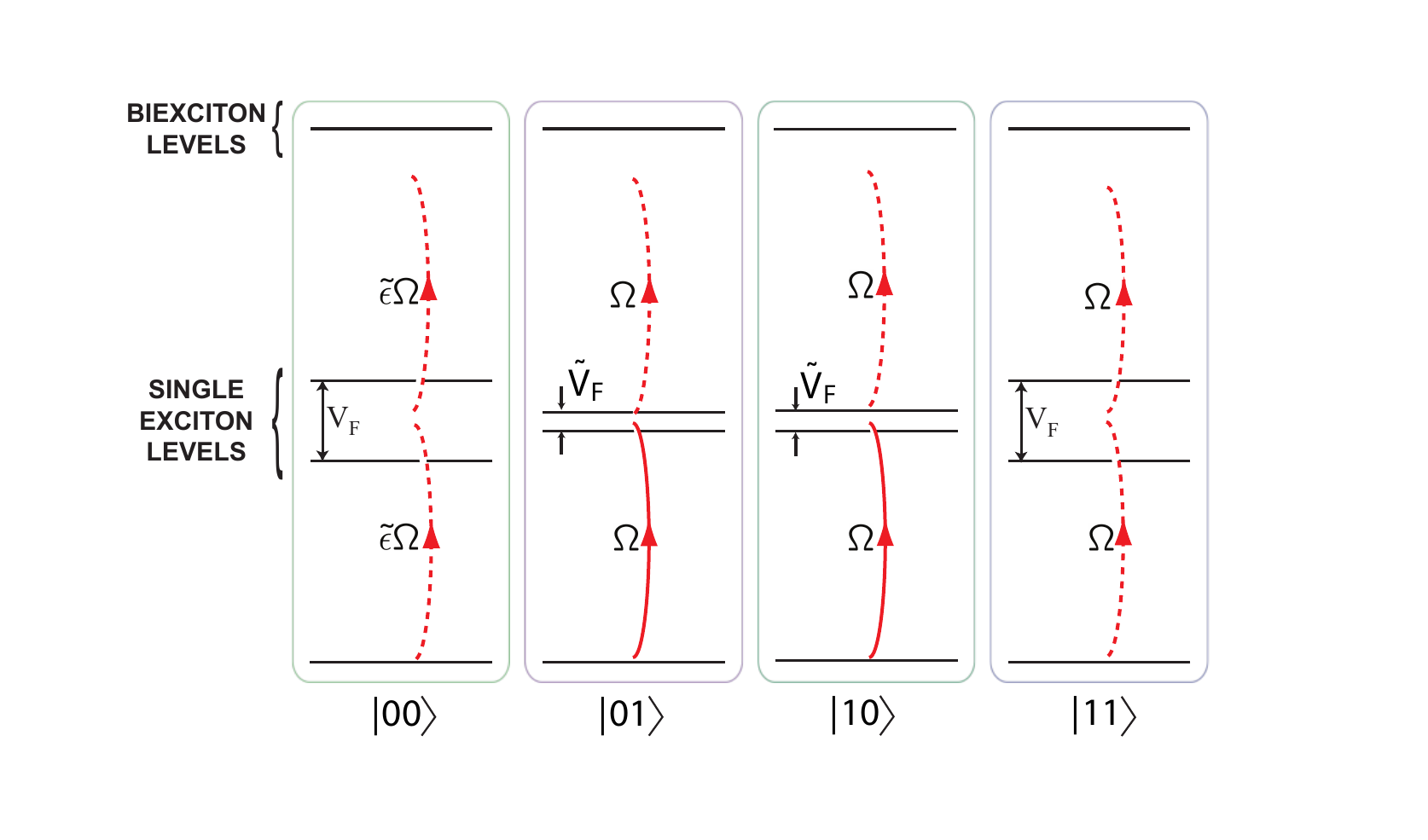}
\caption{The effect of hole-mixing on the level structure of the coupled quantum dot system. The two Foerster interaction terms are $V_{F} = M_{hh,hh}$ and $\tilde{V}_{F} = 2 \tilde{\epsilon} M_{lh,hh}$. Once again, far off resonant transitions are denoted by dashed lines and near resonant transitions by solid lines.}
\label{fig:holemixinglevels}
\end{figure}

With the inclusion of hole-mixing \textit{each} decoupled subspace will now have two single excitonic levels and a biexciton level, and this is illustrated in Fig. \ref{fig:holemixinglevels}. We now discuss whether we can still perform the parity projection in this more complex situation.

First, we must ensure that transitions to excitonic levels in the even subspace remain suppressed. Condition \ref{eq:cond} is still valid to for suppressing transitions from $|11\rangle$. Meanwhile, the couplings to the excitonic levels from the $|00\rangle$ state are reduced by the mixing factor $\epsilon$, and so condition \ref{eq:cond} is also sufficient to suppress these transitions.

Second, transitions within the odd subspace must only occur between the zero and single exciton levels. Our measurement replies on the detection of a single photon: on detection of the photon any population in the biexciton level will be projected into the single excitonic levels, and this must be avoided.

\begin{figure*}[t]
\includegraphics[width = 0.9\textwidth]{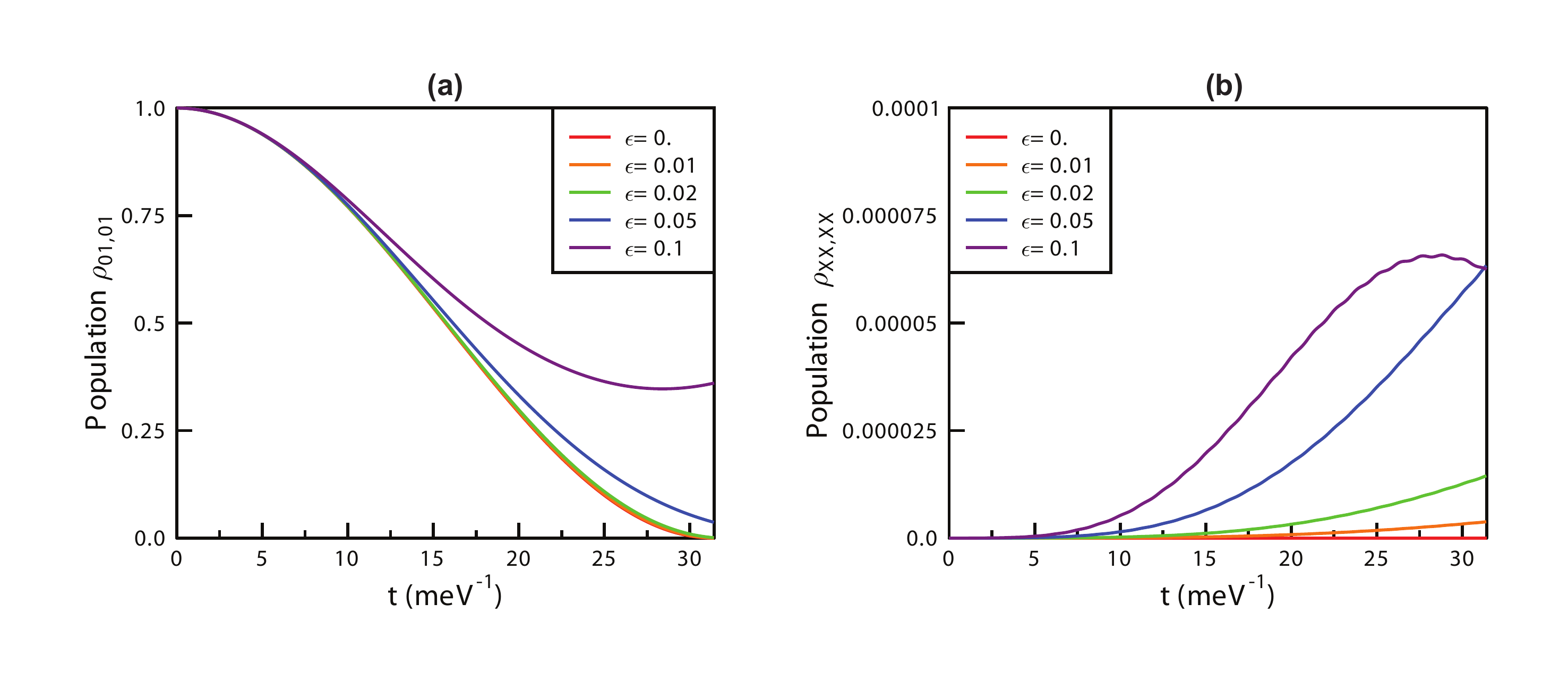}
\caption{Populations of (a) $|01\rangle$ and (b) $|X_{-}X_{+}\rangle$ levels through the duration of the excitation pulse. The parameters are chosen as follows: $M_{hh,hh} = M_{lh,hh} = V_{F}$ and $l_{hh} = l_{lh}$. The laser is tuned to the exciton creation energy $\omega_{0}$ so that $\delta = 0$.}
\label{fig:holemixing}
\end{figure*}

The dynamics is identical for the $|01\rangle$ and $|10\rangle$ subspaces, so we will concentrate only on the levels within the $|01\rangle$ subspace. In the basis $\{|01\rangle, |0X_{-}\rangle, |X_{+}1\rangle, |X_{+}X_{-}\rangle \}$ the Hamiltonian is
\begin{equation}
H = \left( \begin{array}{cccc} 0 & \Omega /2 & \tilde{\epsilon} \Omega /2 & 0 \\ \Omega /2 & \delta & \tilde{V}_{F} & \tilde{\epsilon} \Omega /2 \\ \tilde{\epsilon} \Omega /2 & \tilde{V}_{F} & \delta & \Omega/2 \\ 0 & \tilde{\epsilon} \Omega/2 & \Omega/2 & 2 \delta + V_{XX} \end{array} \right)
\end{equation}
where $\tilde{V}_{F} = 2 \tilde{\epsilon} M_{lh,hh}$. We follow the evolution of the four levels within this subspace over the course of the excitation pulse. We are interested in the populations of the $|01\rangle$ and $|X_{+}X_{-}\rangle$ at the end of the excitation pulse, and so in Fig. \ref{fig:holemixing} we plot the evolution of these populations over time for a range of realistic mixing factors.

We see that, as a result of the strong biexciton shift $V_{XX}$, the population build-up of the $|X_{-}X_{+}\rangle$ is suppressed for a range of realistic mixing factors. For mixing factors upto $\epsilon = 0.05$ the computational states $|01\rangle$ and $|10\rangle$ are also effectively depopulated - and so we conclude that hole mixing not a serious problem for the parity projection.

\subsection{QD Spatial Separation}


\label{sec:spatial}

The two quantum dots that form our CQD structure are naturally spatially separated. As has been mentioned in Section~\ref{sec:scheme}, any source of distinguishability will affect the coherence between the odd-parity states upon detection of a photon. To analyse this effect we derive a new CME describing the detection step, which includes the effects of spatial separation.

This is done in three stages: firstly we derive a Markovian master equation from a microscopic Hamiltonian describing the full dynamics of the system and bath. We then define a jump super-operator which describes the evolution of the system upon a detection event. Finally we are able to identify a CME by imposing the condition that the time averaged CME is equal to the Master Equation derived in the first step.

The Markovian Master equation is derived from the following integro-differential equation:
\begin{equation}
\label{eq:integro-diff}
\dot{\rho}(t) = - \int_{0}^{\infty} d\tau \, Tr_{ph}\{[H_{I}(t),[H_{I}(t-\tau),\rho(t) \otimes \rho_{ph}]]\}
\end{equation}
where $H_{I}(t)$ is the microscopic Hamiltonian in the interaction picture, $\rho(t)$ is the density matrix for the system and $\rho_{ph}$ is the density matrix for the photon bath. If our interaction Hamiltonian is of the general form:
\begin{equation}
\label{eq:factor}
H_{I} = \sum_{i} A_{i}(t) \otimes B_{i}(t)
\end{equation}
then we may write the master equation in the Born-Markov approximation as:
\begin{eqnarray}
\label{eq:me}
\dot{\rho}(t) = - \int_{0}^{\infty} d\tau \, \sum_{\alpha,\beta} C_{\alpha,\beta}(s) [A_{\alpha}^{\dagger}(t) A_{\beta}(t-\tau) \rho(t) \nonumber\\ - A_{\beta}(t-\tau) \rho(t) A_{\alpha}^{\dagger}(t)] + h.c.
\end{eqnarray}
where $C_{\alpha,\beta}(s) = Tr_{ph}[B_{\alpha}^{\dagger}(s) B_{\beta}(0) \rho_{ph}]$ is the enviroment correlation function.

For our case of two quantum dots coupled to a photon bath, the microscopic Hamiltonian is:
\begin{equation}
H = H_{CQD} + H_{ph} + H_{int}
\end{equation}
with
\begin{eqnarray}
H_{CQD} &=& \omega_{0} (c_{X0}^{\dagger} c_{X0} + c_{0X}^{\dagger} c_{0X}) \nonumber \\
H_{ph} &=& \sum_{\mathbf{k}} \omega_{\mathbf{k}} a^{\dagger}_{\mathbf{k}} a_{\mathbf{k}}  \\
H_{int} &=&  \sum_{\mathbf{k}} f(\mathbf{k}) a_{\mathbf{k}} e^{i\mathbf{k}.\mathbf{r}} (c^{\dagger}_{X0} + e^{i  \mathbf{k}.\Delta\mathbf{r}} c^{\dagger}_{0X}) + H. c. \nonumber
\label{eq:Hspatial}
\end{eqnarray}

$c_{X0,0X}$ represents the annihilation operator for an exciton on dot $A,B$ respectively and $a_{\mathbf{k}}$ is the
annihilation operator for a quantum of the electric field. $\mathbf{k}$ is the wavevector for the electric field, $f(\mathbf{k}) = (\mathbf{\hat\mu}.\mathbf{\hat{\sigma}_{\mathbf{k}}}) \epsilon_{\mathbf{k}}$ with $\hat\mu$ the dipole moment vector for each qubit, $\hat{\sigma}_{\mathbf{k}}$ the polarisation vector for the electric field and $\epsilon_{\mathbf{k}}$ is the energy of a mode $\mathbf{k}$ of the electric field. Finally, $\Delta\mathbf{r}$ is the center-to-center separation of the two quantum dots.

We first transform to the interaction picture defined by $H_{0} = H_{CQD} + H_{ph}$:
\begin{equation}
H_{I} = \sum_{\mathbf{k}} f(\mathbf{k}) e^{i \mathbf{k}.\mathbf{r}} e^{i (\omega_{k} - \omega_{0})t} a_{\mathbf{k}} (c^{\dagger}_{X0} + e^{i  \mathbf{k}.\Delta\mathbf{r}} c^{\dagger}_{0X}) + h.c.
\end{equation}
We now proceed to calculate the master equation using expressions Eq. \ref{eq:factor} and Eq. \ref{eq:me}. We must first identify the system and bath operators, $A_{i}(t)$ and $B_{i}(t)$ respectively. We choose:
\begin{eqnarray}
A_{\mathbf{k}}^{\dagger}(t) &=& f(\mathbf{k}) e^{i\mathbf{k}.\mathbf{r}} (c_{X0}^{\dagger} + e^{i\mathbf{k}.\Delta \mathbf{r}} c_{0X}^{\dagger}) e^{i\omega_{0}t} = P_{\mathbf{k}}^{\dagger} e^{i \omega_{0} t}, \nonumber\\
B_{\mathbf{k}}^{\dagger}(t) &=& a_{\mathbf{k}} e^{-i \omega_{\mathbf{k}} t}.
\end{eqnarray}
The master equation then becomes:
\begin{widetext}
\begin{eqnarray}
\dot{\rho}(t) = - 2\pi \sum_{\mathbf{k}} && \frac{1}{2}(1+\tilde{N}(\omega_{\mathbf{k}})) [P_{\mathbf{k}}^{\dagger} P_{\mathbf{k}} \rho(t) - 2 P_{\mathbf{k}} \rho(t) P_{\mathbf{k}}^{\dagger} + \rho(t) P_{\mathbf{k}}^{\dagger} P_{\mathbf{k}}] \delta(\omega_{0}-\omega_{\mathbf{k}}) \nonumber\\ &&+ \frac{1}{2}\tilde{N}(\omega_{\mathbf{k}}) [P_{\mathbf{k}} P_{\mathbf{k}}^{\dagger} \rho(t) - 2 P_{\mathbf{k}}^{\dagger} \rho(t) P_{\mathbf{k}} + \rho(t) P_{\mathbf{k}} P_{\mathbf{k}}^{\dagger}] \delta(\omega_{0}-\omega_{\mathbf{k}}).
\end{eqnarray}
\end{widetext}

As $k_{B}T \ll \omega_{k}$ we may assume that $\tilde{N}(\omega_{\mathbf{k}}) = 0$, and so the master equation back in the Schr\"odinger picture is:
\begin{equation}
\dot{\rho} = -i[H,\rho] + \sum_{\mathbf{k}} (P_{\mathbf{k}} \rho P_{\mathbf{k}}^{\dagger} - \frac{1}{2} \{P_{\mathbf{k}}^{\dagger} P_{\mathbf{k}} \rho + \rho P_{\mathbf{k}}^{\dagger} P_{\mathbf{k}}\}) \delta(\omega_{0} -\omega_{k}).
\end{equation}
Our detection model is simply a single photon detection so we now can define our jump super-operator as:
\begin{equation}
\mathcal{J}[\rho] = \sum_{\mathbf{k'}} P_{\mathbf{k'}} \rho P_{\mathbf{k'}}^{\dagger} \delta(\omega_{0}-\omega_{k})
\end{equation}
where the $k'$-vector runs over the solid angle covered by the detector. It is assumed for simplicity that the detector covers the full solid-angle and also assuming an overall detector inefficiency $\eta$, the stochastic master equation becomes:
\begin{eqnarray}
\dot{\tilde{\rho}} &=& -i[H,\tilde{\rho}] + \sum_{\mathbf{k}} \{(1-\eta) P_{\mathbf{k}} \tilde{\rho} P_{\mathbf{k}}^{\dagger} \nonumber\\ && - \frac{1}{2} \{P_{\mathbf{k}}^{\dagger} P_{\mathbf{k}}, \tilde{\rho}\} \} \delta(\omega_{0}-\omega_{k}).
\end{eqnarray}
Performing the sum over all modes, we obtain
\begin{equation}
\label{cme1}
\dot{\tilde{\rho}} = -i[H,\tilde{\rho}] + (1-\eta ) \mathcal{J}\tilde{\rho} - \mathcal{A}\tilde{\rho}
\end{equation}
where
\begin{eqnarray}
\mathcal{J}\tilde{\rho} &=& \Gamma_{1} [c_{X0} \rho c_{X0}^{\dagger} + 3 f(k_{0}\Delta r) (c_{X0} \rho c_{0X}^{\dagger} + c_{0X} \rho c_{X0}^{\dagger}) \nonumber\\ && + c_{0X} \rho c_{0X}^{\dagger}],\\
\mathcal{A}\tilde{\rho} &=& \Gamma_{1} [c_{X0}^{\dagger} c_{X0} \rho + \rho c_{X0}^{\dagger} c_{0X} + c_{0X}^{\dagger} c_{0X} \rho + \rho c_{0X}^{\dagger} c_{0X}]\nonumber\\
\end{eqnarray}
and
\begin{equation}
f(\alpha) = \frac{2\alpha\cos(\alpha) + (\alpha^2 -2)\sin(\alpha)}{\alpha^3}.
\end{equation}

The function $3 f(k_0 \Delta r)$ characterizes the decohering effect of distinguishable photons. It takes a value of unity for perfectly indistuishable dots, and then Eq.~\ref{cme1} reduces to Eq.~\ref{UNcme}. Stacked self-assembled QDs have separations of the order $5~\text{nm}$~\cite{lovett03b}, while the typical exciton creation energy is $\omega_{0} = 2~\text{eV}$. This gives a value of $3 f(k_{0} \Delta r) = 0.99925$, and we can conclude that spatial separation has a negligible effect on the successful operation of the parity-measurement.

\subsection{Detuning of QD excitonic energy levels}


\label{sec:nonreson}

We have up to this point neglected any inhomogeneity in the underlying structure of the two dots in our coupled system. However, in practice we must expect a certain degree of inhomogeneity: for example, due to the growth technique self-assembled vertically coupled quantum dots will tend to have different sizes. This will result in differences in the confining potentials for the two dots, which in turn will impact on the exciton creation energies, and the overlap integrals which determine the coupling with the laser field. For our parity measurement, the most important effect will come from the detuning or non-resonance of the exciton creation energies for the two dots. This will affect both our ability to perform the excitation step and also the ability to retain coherence when we measure a photon. These effects will be analysed in this section.

\subsubsection{Excitation Pulse}

To begin, let us concentrate on the excitation step. As with the hole-mixing, we have two primary concerns: firstly we must maintain the population of the even-subspace within the computational ground states, and secondly we must ensure that the populations within the odd-subspace are completely transferred to the excitonic levels.

Let us initially focus on the first issue. The Hamiltonian for the $11$ subspace, written in the basis $\{|11\rangle,|1X\rangle,|X1\rangle,|XX\rangle \}$, is now:
\begin{equation}
H = \left( \begin{array}{cccc} 0 & \Omega /2 & \Omega /2 & 0 \\ \Omega /2 & \delta_{B} & V_{F} & \Omega /2 \\ \Omega /2 & V_{F} & \delta_{A} & \Omega/2 \\ 0 & \Omega/2 & \Omega/2 & \delta_{A} + \delta_{B} + V_{XX} \end{array} \right)
\end{equation}
where $\delta_{A,B} = \omega_{A,B} - \omega_{L}$.

As before, we begin by transforming the Hamiltonian into a basis of  $\Omega = 0$ eigenstates. The single exciton subspace is transformed using the following:
\begin{equation}
\left( \begin{array}{c} |1X\rangle \\ |X1\rangle \end{array} \right) = \left( \begin{array}{cc} \cos\theta & -\sin\theta \\ \sin\theta & \cos\theta \end{array} \right) \left( \begin{array}{c} |\psi_{-}\rangle \\ |\psi_{+}\rangle \end{array} \right)
\end{equation}
where $\theta = \arctan(2 V_{F}/(\delta_{A}-\delta_{B}))$ is the mixing angle for the two states $\{|1X\rangle,|X1\rangle\}$, and $\{|\psi_{-}\rangle,|\psi_{+}\rangle\}$ are the new eigenstates. Reintroducing the laser coupling, we find that the Hamiltonian in the new basis $\{|11\rangle,|\psi_{-}\rangle,|\psi_{+}\rangle,|XX\rangle \}$ is:
\begin{equation}
H = \left( \begin{array}{cccc} 0 & \Omega_{-} /2 & \Omega_{+} /2 & 0 \\ \Omega_{-} /2 & \delta^{'}_{B} & 0 & \Omega_{-} /2 \\ \Omega_{+} /2 & 0 & \delta^{'}_{A} & \Omega_{+}/2 \\ 0 & \Omega_{-}/2 & \Omega_{+}/2 & \delta_{A} + \delta_{B} + V_{XX} \end{array} \right)
\end{equation}
where $\Omega_{\pm} = \Omega (\cos\theta \pm \sin\theta)$ and
\begin{eqnarray}
\delta^{'}_{A} &=& \delta_{A} \cos^{2}\theta + \delta_{B} \sin^{2}\theta + V_{F}\sin 2\theta \nonumber \\
\delta^{'}_{B} &=& \delta_{A} \sin^{2}\theta + \delta_{B} \cos^{2}\theta - V_{F}\sin 2\theta
\end{eqnarray}

To suppress any transitions to the excitonic levels, we require that
\begin{equation}
\Omega_{\pm}/2 \ll \delta^{'}_{A},\delta^{'}_{B}, \delta_{A}+\delta_{B}+V_{XX}.
\end{equation}
For both limits of small detuning ($\delta = \delta_{A} - \delta_{B} < 2 V_{F}$) and large detuning ($\delta = \delta_{A} - \delta_{B} > 2 V_{F}$), this condition is satisfied.

Let us now turn our attention to the dynamics within the odd-subspace during the excitation phase. For a successful outcome, we must effect the same population transfer into the exciton levels for both the $|01\rangle$ and $|10\rangle$ initial states. Our Hamiltonian for this subspace in the rotating frame is:
\begin{equation}
H = \left( \begin{array}{cccc} 0 & \Omega /2 & 0 & 0 \\ \Omega /2 & \delta_{A} & 0 & 0 \\ 0 & 0 & 0 & \Omega/2 \\ 0 & 0 & \Omega/2 & \delta_{B}\end{array} \right).
\end{equation}

\begin{figure}[t]
\includegraphics[width = 0.5\textwidth]{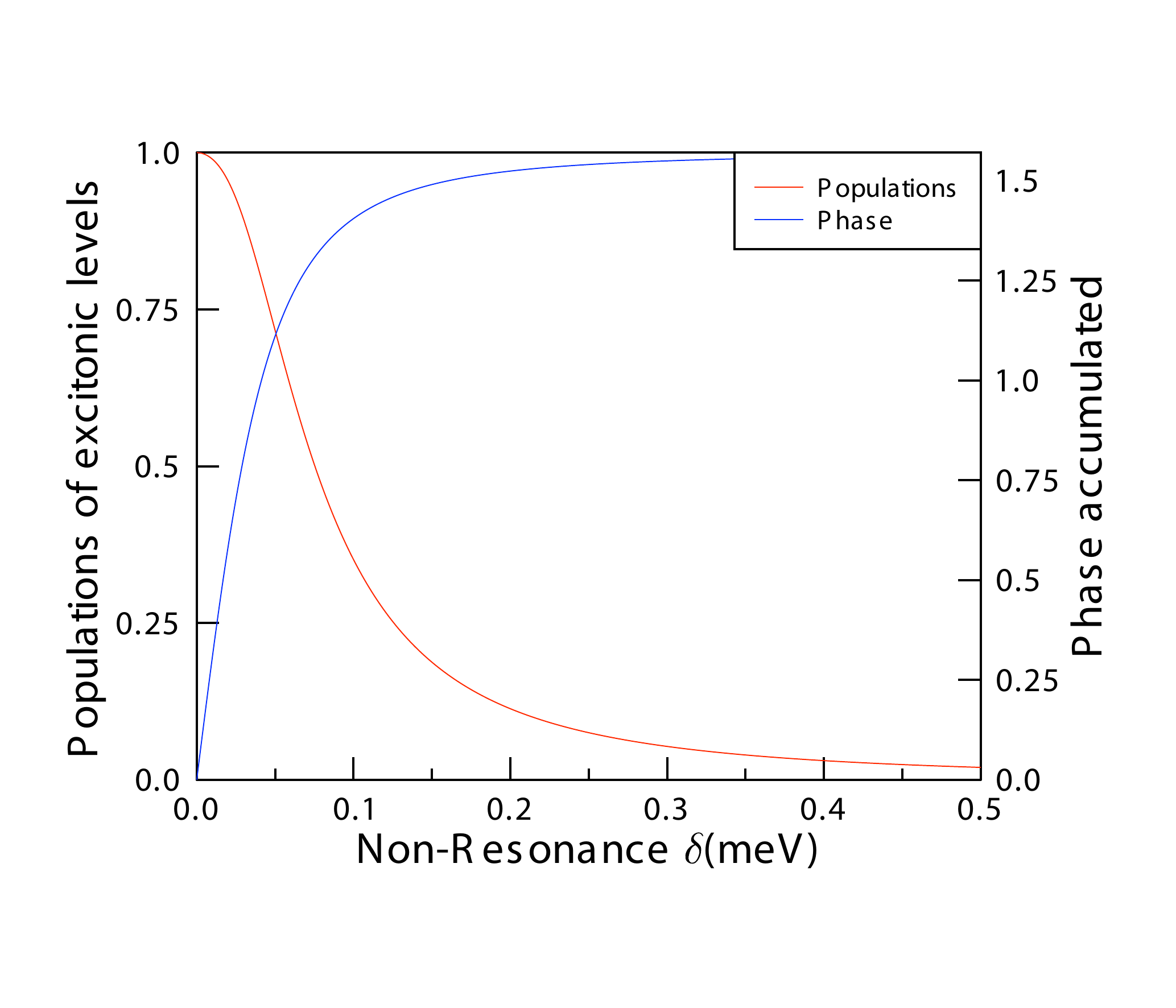}
\caption{Plot of the population of the single excitonic levels and the phase between them at the end of the excitation step as a function of the detuning of the two QDs}
\label{fig:non-reson}
\end{figure}

To achieve equal population transfer to the $|0X\rangle$ and $|X0\rangle$ the laser is tuned to the midpoint between the two excitonic levels: $\delta_{A} = -\delta_{B} = \delta = (\omega_{A} - \omega_{B})/2$. The maximum population of the excitonic levels is given by $\Omega^{2}/(\Omega^{2} + \delta^{2})$ and is found at a time of $\tau_{\pi} = \pi/2\sqrt{\Omega^{2} + \delta^{2}}$. For $\Omega \gg \delta$ we can achieve the required $\pi$ pulse  for each odd-parity state. However, we must also ensure that the coherence between the $|01\rangle$ and $|10\rangle$ states is preserved as we transfer the population to the excited states. Since the levels are not resonant, a net phase difference can be accumulated between them by the end of this excitation pulse. The variation of this phase is shown, along with the population of the excitonic states, in Fig. \ref{fig:non-reson} as a function of detuning. The figure seems to show that we require very similar dots to ensure a successful excitation pulse. However, if the phase accumulated is known, it is possible to correct for at the end of the measurement using single qubit operations.

\vspace{5mm}
\subsubsection{Detection}

We next discuss the effect of detuning on the detection process. To analyse this we will derive a master equation from first principles describing the relaxation process. From here we will proceed to identify the relevant jump operator associated with the detection of a photon, and the resulting CME governing the dynamics of the system until the measurement occurs.

We start with a microscopic Hamiltonian describing the coupled quantum dots, the photon bath, and also the coupling between these two systems:
\begin{equation}
H = H_{CQD} + H_{ph} + H_{int}
\end{equation}
where
\begin{eqnarray}
H_{CQD} &=& \omega_{A} c_{X0}^{\dagger} c_{X0} + \omega_{B}c_{0X}^{\dagger} c_{0X} \nonumber \\
H_{ph} &=& \sum_{\mathbf{k}} \omega_{\mathbf{k}} a^{\dagger}_{\mathbf{k}} a_{\mathbf{k}} \nonumber \\
H_{int} &=& \sum_{\mathbf{k}} g_{\mathbf{k}} a_{\mathbf{k}}^{\dagger}(c_{0X} + c_{X0}) + H. c.
\end{eqnarray}
where $g(\mathbf{k})$ is the photon-exciton coupling constant. Note that in contrast to Eq. \ref{eq:Hspatial} , here we are neglecting spatial separation but accounting for frequency discrepancy.

We begin by transforming into the interaction picture, defined by $H_{0} = H_{CQD} + H_{ph}$. The resulting interaction Hamiltonian is:
\begin{eqnarray}
H_{I} &=& \sum_{\mathbf{k}} g_{\mathbf{k}} a_{\mathbf{k}}^{\dagger} (e^{i (\omega_{\mathbf{k}}-\omega_{A}) t} c_{X0} + e^{i (\omega_{\mathbf{k}}-\omega_{B}) t} c_{0X}) + h.c.\nonumber\\
\end{eqnarray}

We proceed as before (Eqs. \ref{eq:factor} and \ref{eq:me}) by defining the system and environment eigenoperators $A_{i}(t)$ and $B_{i}(t)$, respectively:
\begin{eqnarray}
A_{\mathbf{k}}^{\dagger}(t) &=& g(\mathbf{k}) (e^{i\omega_{A}t} c_{X0}^{\dagger} + e^{i\omega_{B}t} c_{0X}^{\dagger}), \nonumber\\
B_{\mathbf{k}}^{\dagger}(t) &=& a_{\mathbf{k}} e^{-i \omega_{\mathbf{k}} t}.
\end{eqnarray}
Assuming that we are operating at zero temperature, we obtain the following master equation:
\begin{widetext}
\begin{eqnarray}
\dot{\rho}(t) = && \Gamma_{A} c_{X0} \rho(t) c_{X0}^{\dagger} + (\Gamma_{A}+\Gamma_{B})/2 (c_{X0} \rho(t) c_{0X}^{\dagger} e^{i (\omega_{B}-\omega_{A}) t} + c_{0X} \rho(t) c_{X0}^{\dagger} e^{-i (\omega_{B}-\omega_{A}) t}) \nonumber\\ &&+ \Gamma_{B} c_{0X} \rho(t) c_{0X}^{\dagger} - \frac{1}{2} \{c_{X0}^{\dagger} c_{X0} + c_{0X}^{\dagger} c_{0X}, \rho(t)\}
\end{eqnarray}
\end{widetext}
where the decay rates of the two dots are given by $\Gamma_{A,B} = 2\pi \sum_{\mathbf{k}} |g_{\mathbf{k}}|^{2} \delta(\omega_{\mathbf{k}} - \omega_{A,B})$. If ($\omega_{B}-\omega_{A}) \gg (\Gamma_{A}+\Gamma_{B})/2$ the fast oscillating terms may be neglected as their contribution would average to zero on the time scale of the relaxation. In this case, all coherence is lost during the monitored relaxation. Therefore, for a non-destructive measurement, we must work in a regime where ($\omega_{B} - \omega_{A}) \ll (\Gamma_{A}+\Gamma_{B})/2$. This condition is equivalent to requiring that there is a large overlap of the two spectral lines from the two quantum dots, so that the two dots are to a high degree spectrally indistinguishable.

We may make a further assumption to simplify the analysis: the decay rates $\Gamma_{A,B}$ are proportional to $\omega_{A,B}^3$, and so if the detuning is small we may assume that the decay rates are identical for the two dots, ie $\Gamma_{A} = \Gamma_{B} = \Gamma$. The resulting master equation back in the Schr\"odinger picture is then:
\begin{eqnarray}
\dot{\rho}(t) &=& -i [H_{CQD},\rho] + \Gamma (c_{0X} + c_{X0}) ~\rho~ (c_{0X}^{\dagger} + c_{X0}^{\dagger}) \nonumber\\ && - \frac{\Gamma}{2} \{c_{X0}^{\dagger} c_{X0} + c_{0X}^{\dagger} c_{0X}, \rho \}.
\end{eqnarray}

To derive the conditional master equation, we identify the jump operator associated with a photon detection event. As we are assuming that our detector covers the full $4\pi$ solid angle, our jump operator is simply $\mathcal{J}[\rho(t)] = \Gamma (c_{0X} + c_{X0}) ~\rho~ (c_{0X}^{\dagger} + c_{X0}^{\dagger})$. Assuming some loss in the detection process, again parameterised by the variable $\eta$, the unnormalised CME is
\begin{equation}
\dot{\tilde{\rho}} = -i [H_{CQD},\tilde{\rho}] + (1-\eta) \mathcal{J} \tilde{\rho} - \mathcal{A} \tilde{\rho}
\end{equation}
where $\mathcal{A} \tilde{\rho} = \frac{\Gamma}{2} \{c_{X0}^{\dagger} c_{X0} + c_{0X}^{\dagger} c_{0X}, \rho \}$.

We see that when a photon is detected at a time $t_{D}$, a phase of $e^{i (\omega_{A} - \omega_{B}) t_{D}}$ is introduced between the $|01\rangle$ and $|10\rangle$ states. This phase is the relative phase accumulated during the time that the system spends in the excitonic levels.  This is a general problem exhibited by many optical-matter measurement based schemes. Although, in theory we could have information about the relative phase introduced between the two levels, in practice this information can beyond our reach, and effectively introduces a random phase. Such a random phase will destroy any coherence between the two states in the odd-subspace, thereby destroying the non-destructive parity measurement.

In order to retain coherence during the measurement, we therefore require a photon detector with good enough time resolution that we are able to successfully access this phase information. If this is possible, then we may correct for the phase accumulated using single qubit rotations, for example with a single $Z$-rotation on the first QD.

In the worst-case scenario of distinguishable dots, we have a probabilistic scheme for generating entanglement, which is successful only when we project into the even subspace. However, this still provides a powerful resource which may be used for scalable QIP. Furthermore, in this limit of distinguishable dots, the scheme is comparable with many of the existing, inherently probabilistic, schemes for optical-matter measurement-based QIP \cite{barrett05a,lim05}.

\section{Experiments}


\label{sec:verification}

We will conclude by presenting a discussion of experimental procedures for testing the fidelity of the parity measurement.

Quantum process tomography provides a general procedure for characterising the dynamics of a quantum system provided that we can measure each qubit independantly in the $X$, $Y$ and $Z$ bases. However, in our system, we must restrict ourselves to only measuring both qubits in the same basis at the same time, since the two dots are not each individually addressable, and we may only perform global single qubit rotations.


We first introduce entanglement witnesses for the four possible entangled Bell states. It has recently been shown that  entanglement witnesses can be constructed for highly entangled states using the stabilisers that define these states \cite{toth05}. An observable $S_{k}$ is a stabiliser for the state $|\psi\rangle$ if:
\begin{equation}
S_{k} |\psi\rangle = |\psi\rangle .
\end{equation}

The stabilisers for the  four Bell states are then:
\begin{center}
\begin{tabular}{c c}
\hline
Bell States &
Stabilisers $S_{k}$ \\
\hline
$|\psi_{+}\rangle = \frac{1}{\sqrt{2}}|01\rangle + |10\rangle$ &  $-Z_{1}Z_{2}$, $X_{1}X_{2}$  \\
$|\psi_{-}\rangle = \frac{1}{\sqrt{2}}|01\rangle - |10\rangle$ &  $-Z_{1}Z_{2}$, $-X_{1}X_{2}$  \\
$|\phi_{+}\rangle = \frac{1}{\sqrt{2}}|00\rangle + |11\rangle$ &  $Z_{1}Z_{2}$, $X_{1}X_{2}$  \\
$|\phi_{-}\rangle = \frac{1}{\sqrt{2}}|00\rangle - |11\rangle$ &  $Z_{1}Z_{2}$, $-X_{1}X_{2}$  \\
\hline
\end{tabular}
\end{center}

By making measurements on $XX$ and $ZZ$ we are able to distinguish between the four Bell states. The $ZZ$ measurement is related to the parity projection operators by $P_{E,O} = \frac{I \pm Z_{1}Z_{2}}{2}$.  Meanwhile the $XX$ measurement is achieved by first performing a Hadamard gate on each qubit, and then performing the same spin-parity measurement. This $XX$ measurement effectively measures the phase between the states within each subspace. We therefore are using repeated applications of the parity measurement to gain information about sits own operation.

As we have seen from the previous analysis (section \ref{sec:errors}), the primary error source in this system is decoherence between the states in the odd-subspace due to distinguishability of the dots. We can therefore assume that there is some loss of coherence after the parity measurement, as follows:
\begin{eqnarray}
P_{O}\rho P_{O}^{\dagger} &=& P_{01}\rho P_{01}^{\dagger} + \alpha P_{01}\rho P_{10}^{\dagger} + \alpha P_{10}\rho P_{01}^{\dagger} + P_{10}\rho P_{10}^{\dagger} ,\nonumber\\
P_{E}\rho P_{E}^{\dagger} &=& (P_{00} + P_{11})\rho (P_{00}^{\dagger} + P_{11}^{\dagger}). \nonumber\\
\end{eqnarray}
where $P_{i} = |i\rangle \langle i|$. $\alpha$ denotes the degree of coherence; it takes a value of unity for indistinguishable dots, and will be somewhat  less than that for distinguishable dots.

Starting with an initial state that is an equal superposition of the four computational states, the state of the system conditioned on observing a photon is:
\begin{eqnarray}
\rho_{I} = \frac{1}{2}\left( \begin{array}{cccc} 0 & 0 & 0 & 0 \\ 0 & 1 & \alpha & 0 \\ 0 & \alpha & 1 & 0 \\ 0 & 0 & 0 & 0 \end{array} \right).
\end{eqnarray}

Then after a the global Hadamard rotation, the state is:
\begin{eqnarray}
\rho_{II} = \frac{1}{4}\left( \begin{array}{cccc} 1+\alpha & 0 & 0 & -1-\alpha \\ 0 & 1-\alpha & \alpha-1 & 0 \\ 0 & \alpha-1 & 1-\alpha & 0 \\ -1-\alpha & 0 & 0 & 1+\alpha \end{array} \right).
\end{eqnarray}

Finally, the probability of projecting into the odd-subspace after the second parity measurement is $\frac{1-\alpha}{2}$. We therefore see that any loss of coherence during the parity measurements will manifest itself in the final probability, thus giving us a clear method of quantify the effect of distinguishability on the non-destructive nature of the parity measurement.

\section{Conclusions}

In conclusion, we have presented a novel scheme for implementing a spin-parity measurement on a pair of coupled quantum dots. We have estimated the fidelity of the parity measurement scheme presented here in the presence of realistic sources of errors. We find that the measurement is robust in the presence of inefficient detectors, ineffective spin-selective excitation and spatial separation of the dots. For spectrally separated dots, it is found that the performance of the measurement is dependant on the degree of overlap of the spectral lines from the two dots. Total spectral distinguishablility results in a probabilistic measurement which is still sufficient for growing large scale entangled states. Finally, we have proposed an experimental method that is able to verify the success of the parity measurement and quantify the degree to which the measurement can be performed in a non-destructive manner.


\begin{thebibliography}{25}
\expandafter\ifx\csname natexlab\endcsname\relax\def\natexlab#1{#1}\fi
\expandafter\ifx\csname bibnamefont\endcsname\relax
  \def\bibnamefont#1{#1}\fi
\expandafter\ifx\csname bibfnamefont\endcsname\relax
  \def\bibfnamefont#1{#1}\fi
\expandafter\ifx\csname citenamefont\endcsname\relax
  \def\citenamefont#1{#1}\fi
\expandafter\ifx\csname url\endcsname\relax
  \def\url#1{\texttt{#1}}\fi
\expandafter\ifx\csname urlprefix\endcsname\relax\def\urlprefix{URL }\fi
\providecommand{\bibinfo}[2]{#2}
\providecommand{\eprint}[2][]{\url{#2}}

\bibitem[{\citenamefont{Kolli et~al.}()\citenamefont{Kolli, Lovett, Benjamin,
  and Stace}}]{kolli06}
\bibinfo{author}{\bibfnamefont{A.}~\bibnamefont{Kolli}},
  \bibinfo{author}{\bibfnamefont{B.~W.} \bibnamefont{Lovett}},
  \bibinfo{author}{\bibfnamefont{S.~C.} \bibnamefont{Benjamin}},
  \bibnamefont{and} \bibinfo{author}{\bibfnamefont{T.~M.} \bibnamefont{Stace}},
  \bibinfo{note}{quant-ph/0607028}.

\bibitem[{\citenamefont{Kroutvar et~al.}(2004)\citenamefont{Kroutvar, Ducommun,
  Hess, Bichler, Schuh, Abstreiter, and Finlay}}]{kroutvar04}
\bibinfo{author}{\bibfnamefont{M.}~\bibnamefont{Kroutvar}},
  \bibinfo{author}{\bibfnamefont{Y.}~\bibnamefont{Ducommun}},
  \bibinfo{author}{\bibfnamefont{D.}~\bibnamefont{Hess}},
  \bibinfo{author}{\bibfnamefont{M.}~\bibnamefont{Bichler}},
  \bibinfo{author}{\bibfnamefont{D.}~\bibnamefont{Schuh}},
  \bibinfo{author}{\bibfnamefont{G.}~\bibnamefont{Abstreiter}},
  \bibnamefont{and} \bibinfo{author}{\bibfnamefont{J.~J.}
  \bibnamefont{Finlay}}, \bibinfo{journal}{Nature}
  \textbf{\bibinfo{volume}{432}}, \bibinfo{pages}{81} (\bibinfo{year}{2004}).

\bibitem[{\citenamefont{Elzerman et~al.}(2004)\citenamefont{Elzerman, Hanson,
  Bereven, Witkamp, Vandersypen, and Kouwenhoven}}]{elzerman04}
\bibinfo{author}{\bibfnamefont{J.~M.} \bibnamefont{Elzerman}},
  \bibinfo{author}{\bibfnamefont{R.}~\bibnamefont{Hanson}},
  \bibinfo{author}{\bibfnamefont{L.~H. W.~V.} \bibnamefont{Bereven}},
  \bibinfo{author}{\bibfnamefont{B.}~\bibnamefont{Witkamp}},
  \bibinfo{author}{\bibfnamefont{L.~M.} \bibnamefont{Vandersypen}},
  \bibnamefont{and} \bibinfo{author}{\bibfnamefont{L.~P.}
  \bibnamefont{Kouwenhoven}}, \bibinfo{journal}{Nature}
  \textbf{\bibinfo{volume}{430}}, \bibinfo{pages}{431} (\bibinfo{year}{2004}).

\bibitem[{\citenamefont{Petta et~al.}(2005)\citenamefont{Petta, Johnson,
  Taylor, Laird, Yacoby, Lukin, Marcus, Hanson, and Gossard}}]{petta05}
\bibinfo{author}{\bibfnamefont{J.~R.} \bibnamefont{Petta}},
  \bibinfo{author}{\bibfnamefont{A.~C.} \bibnamefont{Johnson}},
  \bibinfo{author}{\bibfnamefont{J.~M.} \bibnamefont{Taylor}},
  \bibinfo{author}{\bibfnamefont{E.~A.} \bibnamefont{Laird}},
  \bibinfo{author}{\bibfnamefont{A.}~\bibnamefont{Yacoby}},
  \bibinfo{author}{\bibfnamefont{M.~D.} \bibnamefont{Lukin}},
  \bibinfo{author}{\bibfnamefont{C.~M.} \bibnamefont{Marcus}},
  \bibinfo{author}{\bibfnamefont{M.~P.} \bibnamefont{Hanson}},
  \bibnamefont{and} \bibinfo{author}{\bibfnamefont{A.~C.}
  \bibnamefont{Gossard}}, \bibinfo{journal}{Science}
  \textbf{\bibinfo{volume}{309}}, \bibinfo{pages}{2180} (\bibinfo{year}{2005}).

\bibitem[{\citenamefont{Loss and DiVicenzo}(1998)}]{loss98}
\bibinfo{author}{\bibfnamefont{D.}~\bibnamefont{Loss}} \bibnamefont{and}
  \bibinfo{author}{\bibfnamefont{D.~P.} \bibnamefont{DiVicenzo}},
  \bibinfo{journal}{Phys. Rev. A} \textbf{\bibinfo{volume}{57}},
  \bibinfo{pages}{57} (\bibinfo{year}{1998}).

\bibitem[{\citenamefont{Knill et~al.}(2001)\citenamefont{Knill, Laflamme, and
  Milburn}}]{knill00}
\bibinfo{author}{\bibfnamefont{E.}~\bibnamefont{Knill}},
  \bibinfo{author}{\bibfnamefont{R.}~\bibnamefont{Laflamme}}, \bibnamefont{and}
  \bibinfo{author}{\bibfnamefont{G.~J.} \bibnamefont{Milburn}},
  \bibinfo{journal}{Nature} \textbf{\bibinfo{volume}{409}}, \bibinfo{pages}{46}
  (\bibinfo{year}{2001}).

\bibitem[{\citenamefont{Lim et~al.}(2005)\citenamefont{Lim, Beige, and
  Kwek}}]{lim05}
\bibinfo{author}{\bibfnamefont{Y.~L.} \bibnamefont{Lim}},
  \bibinfo{author}{\bibfnamefont{A.}~\bibnamefont{Beige}}, \bibnamefont{and}
  \bibinfo{author}{\bibfnamefont{L.~C.} \bibnamefont{Kwek}},
  \bibinfo{journal}{Phys. Rev. Lett.} \textbf{\bibinfo{volume}{95}},
  \bibinfo{pages}{030505} (\bibinfo{year}{2005}).

\bibitem[{\citenamefont{Barrett and Kok}(2006{\natexlab{a}})}]{barrett05a}
\bibinfo{author}{\bibfnamefont{S.~D.} \bibnamefont{Barrett}} \bibnamefont{and}
  \bibinfo{author}{\bibfnamefont{P.}~\bibnamefont{Kok}},
  \bibinfo{journal}{Phys. Rev. A} \textbf{\bibinfo{volume}{71}},
  \bibinfo{pages}{060310} (\bibinfo{year}{2006}{\natexlab{a}}).

\bibitem[{\citenamefont{Terhal and DiVicenzo}(2002)}]{terhal02}
\bibinfo{author}{\bibfnamefont{B.~M.} \bibnamefont{Terhal}} \bibnamefont{and}
  \bibinfo{author}{\bibfnamefont{D.~P.} \bibnamefont{DiVicenzo}},
  \bibinfo{journal}{Phys. Rev. A} \textbf{\bibinfo{volume}{65}},
  \bibinfo{pages}{032325} (\bibinfo{year}{2002}).

\bibitem[{\citenamefont{Beenakker et~al.}(2004)\citenamefont{Beenakker,
  DiVincenzo, Emary, and Kindermann}}]{beenakker04}
\bibinfo{author}{\bibfnamefont{C.~W.~J.} \bibnamefont{Beenakker}},
  \bibinfo{author}{\bibfnamefont{D.~P.} \bibnamefont{DiVincenzo}},
  \bibinfo{author}{\bibfnamefont{C.}~\bibnamefont{Emary}}, \bibnamefont{and}
  \bibinfo{author}{\bibfnamefont{M.}~\bibnamefont{Kindermann}},
  \bibinfo{journal}{Phys. Rev. Lett.} \textbf{\bibinfo{volume}{93}},
  \bibinfo{pages}{020501} (\bibinfo{year}{2004}).

\bibitem[{\citenamefont{Engel and Loss}(2005)}]{engel05}
\bibinfo{author}{\bibfnamefont{H.}~\bibnamefont{Engel}} \bibnamefont{and}
  \bibinfo{author}{\bibfnamefont{D.}~\bibnamefont{Loss}},
  \bibinfo{journal}{Science} \textbf{\bibinfo{volume}{309}},
  \bibinfo{pages}{586} (\bibinfo{year}{2005}).

\bibitem[{\citenamefont{Barrett and Kok}(2006{\natexlab{b}})}]{barrett06b}
\bibinfo{author}{\bibfnamefont{S.~D.} \bibnamefont{Barrett}} \bibnamefont{and}
  \bibinfo{author}{\bibfnamefont{P.}~\bibnamefont{Kok}},
  \bibinfo{journal}{Phys. Rev. B} \textbf{\bibinfo{volume}{73}},
  \bibinfo{pages}{075324} (\bibinfo{year}{2006}{\natexlab{b}}).

\bibitem[{\citenamefont{Laird et~al.}(2006)\citenamefont{Laird, Johnson,
  Marcus, Yacoby, Hanson, and Gossard}}]{laird06}
\bibinfo{author}{\bibfnamefont{E.~A.} \bibnamefont{Laird}},
  \bibinfo{author}{\bibfnamefont{A.~C.} \bibnamefont{Johnson}},
  \bibinfo{author}{\bibfnamefont{C.~M.} \bibnamefont{Marcus}},
  \bibinfo{author}{\bibfnamefont{A.}~\bibnamefont{Yacoby}},
  \bibinfo{author}{\bibfnamefont{M.~P.} \bibnamefont{Hanson}},
  \bibnamefont{and} \bibinfo{author}{\bibfnamefont{A.~C.}
  \bibnamefont{Gossard}}, \bibinfo{journal}{Phys. Rev. Lett.}
  \textbf{\bibinfo{volume}{97}}, \bibinfo{pages}{056801}
  (\bibinfo{year}{2006}).

\bibitem[{\citenamefont{Lovett et~al.}(2005)\citenamefont{Lovett, Nazir, Pazy,
  Barrett, Spiller, and Briggs}}]{lovett05}
\bibinfo{author}{\bibfnamefont{B.~W.} \bibnamefont{Lovett}},
  \bibinfo{author}{\bibfnamefont{A.}~\bibnamefont{Nazir}},
  \bibinfo{author}{\bibfnamefont{E.}~\bibnamefont{Pazy}},
  \bibinfo{author}{\bibfnamefont{S.~D.} \bibnamefont{Barrett}},
  \bibinfo{author}{\bibfnamefont{T.~P.} \bibnamefont{Spiller}},
  \bibnamefont{and} \bibinfo{author}{\bibfnamefont{G.~A.~D.}
  \bibnamefont{Briggs}}, \bibinfo{journal}{Phys. Rev. B}
  \textbf{\bibinfo{volume}{72}}, \bibinfo{pages}{115324}
  (\bibinfo{year}{2005}).

\bibitem[{\citenamefont{Wiseman}(1994)}]{wiseman94}
\bibinfo{author}{\bibfnamefont{H.}~\bibnamefont{Wiseman}}, Ph.D. thesis,
  \bibinfo{school}{University of Queensland} (\bibinfo{year}{1994}).

\bibitem[{\citenamefont{Gardiner and Zoller}(2000)}]{gardiner00}
\bibinfo{author}{\bibfnamefont{C.~W.} \bibnamefont{Gardiner}} \bibnamefont{and}
  \bibinfo{author}{\bibfnamefont{P.}~\bibnamefont{Zoller}},
  \emph{\bibinfo{title}{Quantum {N}oise}} (\bibinfo{publisher}{Springer},
  \bibinfo{year}{2000}).

\bibitem[{\citenamefont{Stace et~al.}(2003)\citenamefont{Stace, Milburn, and
  Barnes}}]{stace03}
\bibinfo{author}{\bibfnamefont{T.~M.} \bibnamefont{Stace}},
  \bibinfo{author}{\bibfnamefont{G.~J.} \bibnamefont{Milburn}},
  \bibnamefont{and} \bibinfo{author}{\bibfnamefont{C.~H.~W.}
  \bibnamefont{Barnes}}, \bibinfo{journal}{Phys. Rev. B}
  \textbf{\bibinfo{volume}{67}}, \bibinfo{pages}{085317}
  (\bibinfo{year}{2003}).

\bibitem[{\citenamefont{Birkedal et~al.}(2001)\citenamefont{Birkedal, Leosson,
  and Hvam}}]{birkedal01}
\bibinfo{author}{\bibfnamefont{D.}~\bibnamefont{Birkedal}},
  \bibinfo{author}{\bibfnamefont{K.}~\bibnamefont{Leosson}}, \bibnamefont{and}
  \bibinfo{author}{\bibfnamefont{J.~M.} \bibnamefont{Hvam}},
  \bibinfo{journal}{Phys. Rev. Lett.} \textbf{\bibinfo{volume}{87}},
  \bibinfo{pages}{227401} (\bibinfo{year}{2001}).

\bibitem[{\citenamefont{Biolatti et~al.}(2002)\citenamefont{Biolatti, D'Amico,
  Zanardi, and Rossi}}]{biolatti02}
\bibinfo{author}{\bibfnamefont{E.}~\bibnamefont{Biolatti}},
  \bibinfo{author}{\bibfnamefont{I.}~\bibnamefont{D'Amico}},
  \bibinfo{author}{\bibfnamefont{P.}~\bibnamefont{Zanardi}}, \bibnamefont{and}
  \bibinfo{author}{\bibfnamefont{F.}~\bibnamefont{Rossi}},
  \bibinfo{journal}{Phys. Rev. B} \textbf{\bibinfo{volume}{65}},
  \bibinfo{pages}{075306} (\bibinfo{year}{2002}).

\bibitem[{\citenamefont{Lovett et~al.}(2003)\citenamefont{Lovett, Reina, Nazir,
  and Briggs}}]{lovett03b}
\bibinfo{author}{\bibfnamefont{B.~W.} \bibnamefont{Lovett}},
  \bibinfo{author}{\bibfnamefont{J.~H.} \bibnamefont{Reina}},
  \bibinfo{author}{\bibfnamefont{A.}~\bibnamefont{Nazir}}, \bibnamefont{and}
  \bibinfo{author}{\bibfnamefont{G.~A.~D.} \bibnamefont{Briggs}},
  \bibinfo{journal}{Phys. Rev. B} \textbf{\bibinfo{volume}{68}},
  \bibinfo{pages}{205319} (\bibinfo{year}{2003}).

\bibitem[{\citenamefont{Borri et~al.}(2001)\citenamefont{Borri, Langbein,
  Schneider, Woggon, Sellin, Ouyang, and Bimberg}}]{borri01}
\bibinfo{author}{\bibfnamefont{P.}~\bibnamefont{Borri}},
  \bibinfo{author}{\bibfnamefont{W.}~\bibnamefont{Langbein}},
  \bibinfo{author}{\bibfnamefont{S.}~\bibnamefont{Schneider}},
  \bibinfo{author}{\bibfnamefont{U.}~\bibnamefont{Woggon}},
  \bibinfo{author}{\bibfnamefont{R.~L.} \bibnamefont{Sellin}},
  \bibinfo{author}{\bibfnamefont{D.}~\bibnamefont{Ouyang}}, \bibnamefont{and}
  \bibinfo{author}{\bibfnamefont{D.}~\bibnamefont{Bimberg}},
  \bibinfo{journal}{Phys. Rev. Lett.} \textbf{\bibinfo{volume}{87}},
  \bibinfo{pages}{157401} (\bibinfo{year}{2001}).

\bibitem[{\citenamefont{Nielsen and Chuang}(2000)}]{nielsen00}
\bibinfo{author}{\bibfnamefont{M.~A.} \bibnamefont{Nielsen}} \bibnamefont{and}
  \bibinfo{author}{\bibfnamefont{I.~L.} \bibnamefont{Chuang}},
  \emph{\bibinfo{title}{Quantum Computation and Quantum Information}}
  (\bibinfo{publisher}{Cambridge}, \bibinfo{year}{2000}).

\bibitem[{\citenamefont{Bayer et~al.}(1999)\citenamefont{Bayer, Kuther,
  Forchel, Gorbunov, Timofeev, Schafer, Reithmaier, Reinecke, and
  Walck}}]{bayer99}
\bibinfo{author}{\bibfnamefont{M.}~\bibnamefont{Bayer}},
  \bibinfo{author}{\bibfnamefont{A.}~\bibnamefont{Kuther}},
  \bibinfo{author}{\bibfnamefont{A.}~\bibnamefont{Forchel}},
  \bibinfo{author}{\bibfnamefont{A.}~\bibnamefont{Gorbunov}},
  \bibinfo{author}{\bibfnamefont{V.~B.} \bibnamefont{Timofeev}},
  \bibinfo{author}{\bibfnamefont{F.}~\bibnamefont{Schafer}},
  \bibinfo{author}{\bibfnamefont{J.~P.} \bibnamefont{Reithmaier}},
  \bibinfo{author}{\bibfnamefont{T.~L.} \bibnamefont{Reinecke}},
  \bibnamefont{and} \bibinfo{author}{\bibfnamefont{S.~N.} \bibnamefont{Walck}},
  \bibinfo{journal}{Phys. Rev. Lett.} \textbf{\bibinfo{volume}{82}},
  \bibinfo{pages}{1748} (\bibinfo{year}{1999}).

\bibitem[{\citenamefont{Luttinger and Kohn}(1955)}]{luttinger55}
\bibinfo{author}{\bibfnamefont{J.~M.} \bibnamefont{Luttinger}}
  \bibnamefont{and} \bibinfo{author}{\bibfnamefont{W.}~\bibnamefont{Kohn}},
  \bibinfo{journal}{Physical Review} \textbf{\bibinfo{volume}{97}},
  \bibinfo{pages}{869} (\bibinfo{year}{1955}).

\bibitem[{\citenamefont{Toth and Guhne}(2005)}]{toth05}
\bibinfo{author}{\bibfnamefont{G.}~\bibnamefont{Toth}} \bibnamefont{and}
  \bibinfo{author}{\bibfnamefont{O.}~\bibnamefont{Guhne}},
  \bibinfo{journal}{Phys. Rev. A} \textbf{\bibinfo{volume}{72}},
  \bibinfo{pages}{022340} (\bibinfo{year}{2005}).

\end{thebibliography}
\end{document}